\documentstyle[aps,fleqn,tighten,epsfig,preprint]{revtex}
\begin{document}
\draft
\title{Comprehensive Evidences of  Octupole Vibration in  $^{158}$Gd}
\author{G.L.Long$^{1,2,3}$\thanks{Correspondence address: Department of
Physics,
Tsinghua University, Beijing 100084, P.R. China; email:
gllong@mail.tsinghua.edu.cn}
,H.Y.Ji$^1$ ,E. G. Zhao$^{1,2}$}
\address{
$^{1}$Department of Physics, Tsinghua University, Beijing 
100084, China\\
$^2$Institute of Theoretical Physics, Chinese Academy of 
Sciences, Beijing, China\\
$^3$Center of Theoretical Nuclear Physics, National Laboratory\\
 of Heavy Ion Accelerator, Chinese Academy of Sciences, 
 Lanzhou, 730000,P R China}
\date{Feb. 1, 1999}
\maketitle

\begin{abstract}
Comprehensive evidences of the SU(3) limit in the spdf 
interacting boson model, 
a dynamical symmetry describing octupole vibration in 
rotatonal nucleus,  are found in the spectrum,
E2 and E1 transition rates, and relative intensities 
in $^{158}$Gd. This gives a good
example of rotational nucleus with octupole vibration in rare-earth region.
\end{abstract}


There have been continued interests in the studies of octupole degree of
freedom in nuclear structure recently
\cite{r1,r2,r3,r4,r5,r6,r7,r8,r9,r9a,r10,r11,r12,r13,r14,r15,r16,r16a,r17,r18,r19,r20,r20p}. 
In the boson model, negative parity
states are described by the spdf 
interacting boson model(IBM)\cite{r9,r9a,r10,r11,r12,r13,r14,r15,r16,r17} 
or the sdf 
IBM\cite{r18,r19}.
Gamma soft octupole deformation has been found in Ba isotopes\cite{r15}. 
As for rotation with octupole deformation, corresponding to the SU(3) limit
in the spdf IBM, 
the experimental evidence has not yet been found. 
It has been pointed that $^{232}$U and other actinide nuclei 
may be candidates for the SU(3) limit\cite{r11,r12,r13,r17}. However, 
while the spectrum agrees with  theoretical calculation very 
well, there are  few electromagnetic transition data,
in particular E1 transitions connecting  positive and negative parity states. 
Experimental evidences of nuclei with such a dynamical symmetry
is still lacking. Whereas
these dynamic symmetries are very important because they can be 
used to classify states and characterize collective features of a nucleus. 
To this purpose we have studied the structures of over 30 deformed 
nuclei in the rare earth and actinide region, and found 
comprehensive evidences of octupole vibration in $^{158}$Gd.
This result is somewhat out of the general expectation that a 
good example nucleus of the SU(3) octupole deformation dynamical 
symmetry should be in the actinide region\cite{r11,r12,r13,r17}.

The octupole vibration in rotational nucleus is characterized by 
the SU(3) group chain,
\begin{eqnarray}
&&U(16)\supset U(6)\otimes U(10)\supset SU_{sd}(3)\otimes SU_{pf}(3)
\supset SU_{spdf}(3)\supset O(3)\nonumber\\
&& \;\;\;N\;\;\;\;\;\;\;\;\;\;N_+\;\;\;\;\;\;\;N_-\;\;\;\;\;\;(\lambda_+
\mu_+)\;\;\;\;\;\;(\lambda_- \mu_-)\;\;\;\;\;\;\;\;(\lambda \mu)\;\;\;\;
\;\;\;\;\;\;\;L,
\end{eqnarray}
and the energy eigenvalue is
\begin{eqnarray}
E=\epsilon_-N_{pf}+a_1C_{2SU_+}+a_2C_{2SU_-}+a_3C_{2SU(3)}+a_4L(L+1).
\end{eqnarray}

 The g.s.-band,$\beta$-band and $\gamma$-band are generated from
(2N,0),(2N-4,2)K=0 and (2N-4,2)K=2 respectively,with N sd-bosons, 
where N is the 
valence nucleon pair number. In $^{158}$Gd, N is equal to 13. 
The low-lying negative parity are generalized by the SU(3) 
irreducible representation(IR) from the decomposition of 
(2N-2,0)$\otimes$(3,0); that is $K^p=0^-$ from (2N+1,0)
 and $K^p=1^-$ from (2N-1,1) respectively. 
The value of $a_2$ is taken zero, since it 
is irrelevant to the spectrum of the low-lying 
states with only one pf-boson. The parameters are 
then determined by experimental data.
They are: $a_1$=-7.793keV,$\epsilon_-$=3.160MeV,
$a_3$=-4.686keV. And $a_4$ is 11.917keV for the 
positive parity states and 8.592keV for the negative 
parity states, respectively. The smaller value of $a_4$ 
for the negative parity states reflects an increase of moment 
of inertia for the negative parity states due to octupole deformation, 
an effect which has also been observed in Uranium isotopes\cite{r21}. 
The spectrum of the spdf SU(3) is compared with data\cite{r19,r22,r22a} 
in Fig.1. 
The general agreement between experiment and calculation is good. 
The five low-lying bands, 3 with positive parity and 2 with negative parity, 
are all well reproduced. But from the spectrum alone, it is not sufficient
to determine the nature of the dynamical symmetry. The more tough criteria
in determining the nature of the collective motion lie in the 
electromagnetic transition part.

E2 transitions among the positive parity states are calculated 
using the transition operator: $T(E2)^2=e_2((s^{\dagger}\tilde{d}
+d^{\dagger}s)^2-{\sqrt{7}\over 2}
 (d^{\dagger}\tilde{d})^2)$, the SU(3) generator. The calculated 
B(E2) values are compared with experimental data in table \ref{t1}. 
The agreement with experimental data is very good. The inband transitions
in the ground state band agree with the data well. 
Since the SU(3)generator does not have matrix elements 
between different SU(3) IR's, all the
inter-band E2 transitions from $\gamma$ or $\beta$ bands 
to the ground state band are zero.
This is in agreement with the experimental data. All the 
interband transitions are very weak.
A small breaking in the SU(3) dynamical symmetry will 
produce nonzero interband transitions.
This is a second order effect, and in this work, we are 
not going to pursue the details.
  At high spins($L=10$), the theoretical B(E2) is less 
than the data. This is the well-known reduction of 
collectivity problem in boson models and can only be 
solved by considering the g-boson\cite{r23}.

There are also ample experimental data on the electric 
dipole transitions. We have calculated the E1 transitions 
using standard group theoretic method as in Ref.\cite{r24}. 
The E1 transition operator is taken the following form: 
$T(E1)=e ((s^{\dagger}\tilde{p}+p^{\dagger})^1+\chi_{dp}
(d^{\dagger}\tilde{p}+p^{\dagger}\tilde{p})^1+
\chi_{df}(d^{\dagger}\tilde{f}+f^{\dagger}\tilde{d})^1)$. 
These parameters are determined by experimental data. The values are :
$\chi_{dp}=-3.825$, $\chi_{df}=3.676$. The result of this parametriztion
is labeled as Cal1 in table \ref{t2}. The agreement between calculation 
and data is very well.  
The transitions from $0^-$ band to the ground state band transitions are
perfectly reproduced. In particular, the transition from $2^-$ to 
$2_{\beta}$ is much less than the transition from $1^-_1$ to ground 
state in experiment, and this is also reproduced by the calculation 
well. Considering the large range of variations in the data, the agreement
is remarkable. 
It is worth pointing out that the transition operator is quite close the
generator of O(10)\cite{r13}, where $\chi_{dp}=-1.2649$ and $\chi_{df}=1.1832$. 
This also shows the importance of the p boson 
 in describing octupole collective motions, which has been pointed by
many
 authors\cite{r10,r11,r12,r15,r16,r16a}. In order to see the goodness
 of the generator form,
we made a calculation for the E1 transitions using just generator, 
the results are also listed in table \ref{t2} labeled as Cal2. It is
 apparent that the  generator  has already given a satisfactory 
agreement with  the data. 
We have also calculated the relative intensities. The results using the
E1 transition operator determined from experiment  are 
listed in table \ref{t3}. We see that the agreement between calculation 
and data is very well. We have also calculated the relative intensities
using the O(10) generator, whose results are not shown here. The
agreement between calculation and the data is quite good.

From these comparisons of the spectrum, the E2 and E1 transition rates,
 and relative intensities, we can conclude that $^{158}$Gd is a good 
example of the SU(3) dynamical symmetry in the spdf IBM, a dynamical
 symmetry describing the rotations with octupole vibration. 
In particular, all  existing electric dipole transition data varying 
over a large range agree with the SU(3) limit dynamical symmetry 
results very well. This has passed 
through the stringent test on the validity of the dynamical symmetry, 
that is, the check on the wave functions  of a dynamical symmetry. 
This has established firmly the experimental evidence of  rotation 
with octupole vibration in $^{158}$Gd. Because of the similarities 
of the Gd isotopes with other rare-earth nuclei in many properties,
it is hoped that this finding will be helpful in the studies of the 
octupole collectivity in this region.

The authors thank the financial support of the China National
 Natural Science Foundation, Fok Ying Tung Education foundation, 
Excellent University Young Faculty Fund of China Education Ministry.
 Helpful discussions with Prof. Hongzhou Sun and Qizhi Han are also 
gratefully acknowledged.

\begin{table}
\begin{center}
\begin{tabular}{llll|llll}\hline
$I_i$ & $I_f$ & Exp  &\multicolumn{1}{l}{ Cal} &$I_i$ & $I_f$ 
& Exp  & Cal\\ \hline
\multicolumn{4}{l|}{Intra band transitions} & \multicolumn{4}{l}
{Interband transitions} \\
$2_g$ & $0_g$ & 198(6) & 198 & $2_{\beta}$ & $4_g$ & 1.39(15) & 0.0\\
$4_g$ & $2_g$ & 289(5) & 279 & $2_{\beta}$ & $0_g$ & 0.31(4)  & 0.0\\
$6_g$ & $4_g$ &        & 300 & $2_{\gamma}$ & $4_g$ & 0.27(4) & 0.0\\
$8_g$ & $6_g$ & 320(30) & 302 &$4_{\gamma}$ & $2_g$ & $5.9(7)$ & 0.0\\
$10_g$ & $8_g$ & 330(30) & 296 &$2_{\gamma}$ & $0_g$ & 3.5(4) & 0.0\\
$12_g$& $10_g$ & 310(30) & 282 &             &       &        &\\
\hline
\end{tabular}
\end{center}
\caption{B(E2) values among the positive parity states.}
\label{t1}
\end{table}

\begin{table}
\begin{center}
\begin{tabular}{cllll}\hline
   $I_i$ &   $ I_f$   &   Exp.(W.u.)  & Cal.1(W.u.) &  Cal.2(W.u.)\\ \hline
  $1^-_2$&   $0^+_1$  &   0.0035(8)   & 0.0028      &  0.0035     \\
  $1^-_2$&   $2^+_1$  &   0.0063(16)   & 0.0056      &  0.0068     \\
  $3^-_1$&   $2^+_1$  &   0.00033(10)  & 0.00029      &  0.00047     \\
  $3^-_1$&   $4^+_1$  &   0.00029(8)  & 0.00041      &  0.00088     \\
  $2^+_{\beta}$&   $1^-_1$  &   $6.4(8)\times 10^{-5}$ & $1.30\times 
10^{-5}$ &  $6.2\times 10^{-5}$   \\
  $2^+_{\beta}$&   $2^-_1$  & $1.21(5)\times10^{-5}$ & $1.19\times
10^{-5}$& $3.16 \times10^{-5}$      \\
  $2^+_{\beta}$&   $3^-_1$  &  $1.89(24)\times 10^{-4}$ & $1.11\times
 10^{-5}$  & $2.54\times 10^{-4}$ \\\hline
\end{tabular}
\end{center}
\caption{Comparison of B(E1) values in $^{158}$Gd.}
\label{t2}
\end{table}
\begin{table}
\begin{center}
\doublerulesep 1.5pt
\begin{tabular}{c l l l l l l }\hline\hline
  Nucleus    &  $E_{level}(keV)$  & $K^{\pi}$  & $I_i$  &   $ I_f$ 
&   Cal.  &   Exp.   \\\hline
  $^{158}$Gd &  977               &  $1^-$     &$1^-_1$ &   $0^+_{gs}$
&   100   &  100(5)  \\\
             &                    &            &        &   $2^+_{gs}$
&  51     &  76(4)   \\\
             &  1042              &            &$3^-_1$ &   $2^+_{gs}$
&  100    &  100(20) \\\
             &                    &            &        &   $4^+_{gs}$
&   76    &  47.2(9) \\\
             &  1176              &            &$5^-_1$ &   $4^+_{gs}$
&  100    &  100(6)  \\\
             &                    &            &        &   $6^+_{gs}$
& 75.9    &  26.7(16)\\\
             &  1260              &         &$2^+_{\beta}$ &$1^-_1$
&  5.1    &  5.1(46) \\\
             &                    &            &        &   $2^-_1$
&  2.75   &  0.56(4) \\\
             &                    &            &        &   $3^-_1$
&  2.01   &  6.9(6)  \\\
             &  1263              &  $0^-  $   &$1^-_2$ &   $0^+_{gs}$
&  64     &  68(4)   \\\
             &                    &            &        &   $2^+_{gs}$
&  100    &  100(6)  \\\
             &  1403              &            &$3^-_2$ &   $2^+_{gs}$
&  100    &  100(6)  \\\
             &                    &            &        &   $4^+_{gs}$
&  85     &  85(5)   \\\
             &                    &            &        &   $2^+_{\beta}$
& 0.002   &  0.04(1) \\\
             &  1407              &  $1^-$   &$4^+_{\beta}$ &$3^-_1$
&  19.9   &  19.9(12)\\\
             &                    &            &        &   $4^-_1$
&  1.29   &  0.33(2) \\\
             &                    &            &        &   $5^-_1$
&  5.21   &  6.6(5)  \\\
             &  1639              &  $0^-  $   &$5^-_2$ &   $4^+_{gs}$
&  100    & 100(8)   \\\
             &                    &            &        &   $6^+_{gs}$
&  61     & 33(5)     \\\hline
\end{tabular}
\end{center}
\caption{Comparisons of relative intensities in $^{158}$Gd}
\label{t3}
\end{table}
\begin{figure}
\begin{center}
\epsfig{figure=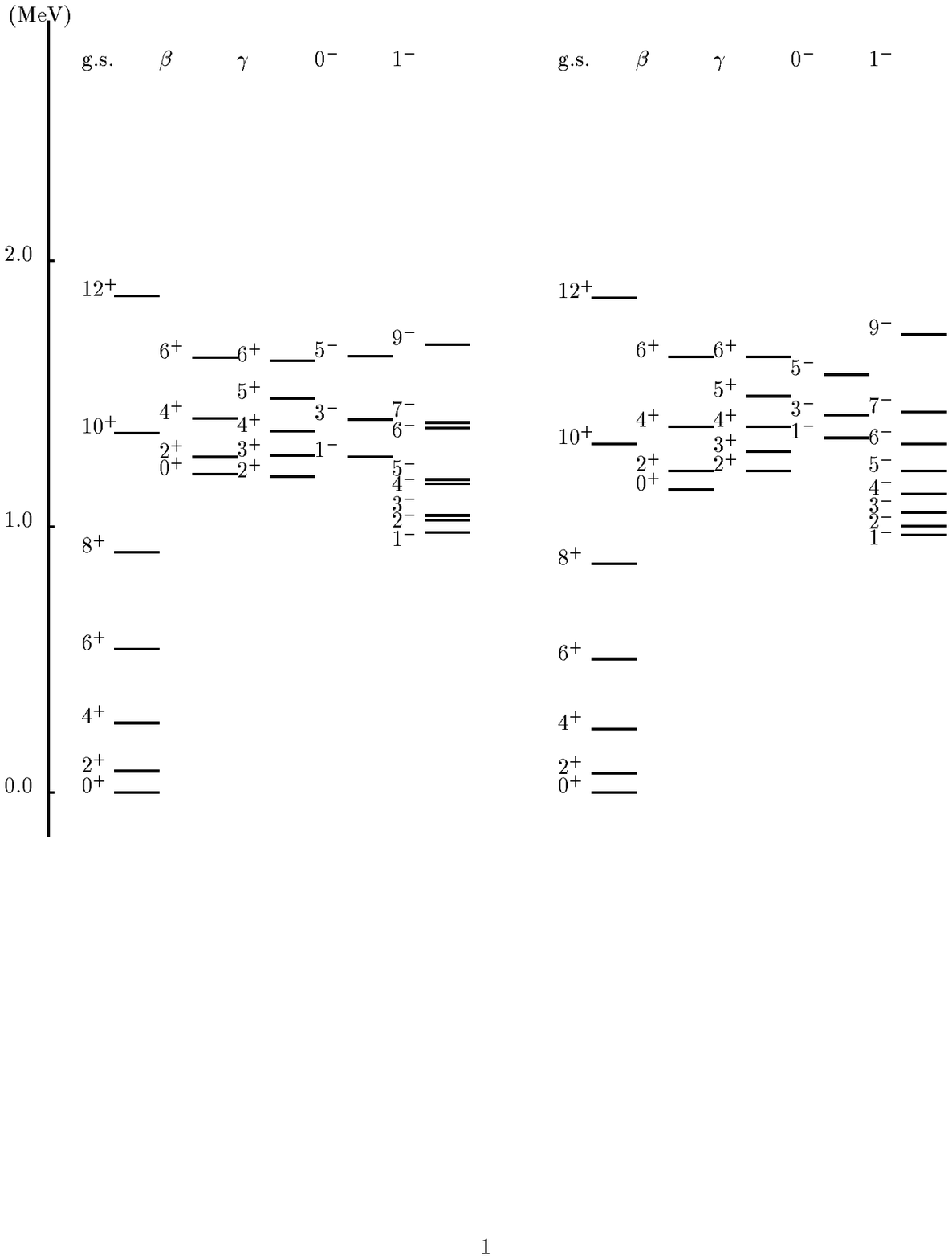,width=13cm}
\caption{The spectra of $^{158}$Gd. The left part is the experimental
spectrum, and the right part is the calculated spectrum.}
\end{center}
\end{figure}
\end{document}